\documentclass[mathleft
]{an}
\usepackage{aas_macros}
\usepackage[sort,authoryear]{natbib}
\usepackage{graphicx}
\usepackage{times}
\bibpunct{(}{)}{;}{a}{}{,}
\begin{document}

\Pagespan{1}{6}
\Yearpublication{2013}%
\Yearsubmission{2012}%
\Month{08}%
\Volume{999}%
\Issue{99}%

\title{Acoustic and buoyancy modes throughout stellar
  evolution -- \\
  Seismic properties of stars at different stellar ages and
  masses}

\author{S. Schuh\inst{1}\fnmsep\thanks{Corresponding author:
  \email{schuh@astro.physik.uni-goettingen.de}\newline}
}
\titlerunning{Acoustic and buoyancy modes throughout stellar evolution}
\authorrunning{S. Schuh}
\institute{
  Georg-August-Universit\"at G\"ottingen, Institut f\"ur Astrophysik,
  Friedrich-Hund-Platz~1, 37077 G\"ottingen, Germany
}

\received{5 September 2012}
\accepted{22 October 2012}
\publonline{later}


\abstract{%
Parameter regions in which stars can become pulsationally
unstable are found throughout the Hertzsprung-Russel
diagram. Stars of high, intermediate, low and very low masses
may cross various instability regions along their paths of
evolutionary sequences. In describing them, I give special
consideration to hybrid pulsational characteristics that are
particularly valuable for asteroseismic investigations, to
\smash{$\dot{P}$} measurements that allow us to directly
follow the stellar evolutionary changes in some stars, and
to new research results that stand out with respect to
previous consensus.}
\maketitle
\section{Introduction}
Stars are gaseous spheres in hydrostatic equilibrium that
continuously radiate away energy into space. The hydrostatic
equilibrium, when disturbed, can be balanced again by
adjusting the pressure; local deviations from a spherical
shape can be restored by the action of buoyancy. The energy
carried from the centre to the surface of the star can under
the right circumstances be temporarily stored and
transformed into kinetic energy, giving rise to oscillations
about the equilibrium. The properties of these stellar
pulsations can vary widely, and many possible combinations
of circumstances that are 'right' cause pulsating stars to
appear all over the Hertzsprung-Russel diagram. A visual
representation of many of these classes of pulsators has
been updated several times from Fig.~{1} in
\citet{1998Ap&SS.261....1C}. A very similar and improved
recent representation may be found as Fig.~{1 (right panel)}
in the comprehensive and at the same time readily
comprehensible summary of the subject of asteroseismology by
\citet{2012arXiv1205.6407H}.
\par
I will discuss a cross-section of these pulsating stars and
their seismic properties for different masses and
ages. Since it depends on the mass how a star evolves with
age, mass ranges are distinguished from the point of view of
different end products of stellar evolution: by high-mass
stars I mean those with initial masses on the main sequence
above eight solar masses, typically bound to end in a
supernova explosion. There is a possible exception for stars
around eight to ten solar masses that can go through carbon
fusion only ('super-Asymptotic Giant Branch') before evolving
into \ion{O}{}/\ion{Mg}{}/\ion{Ne}{} core White
Dwarfs. I designate as intermediate- and low-mass stars
those with initial masses between $0.5$ and eight solar
masses that will evolve into \ion{C}{}/\ion{O}{} core
White Dwarfs via a central helium burning stage.  Low-mass
stars have masses between $0.08$ and $0.5$ solar masses and
will eventually turn into helium-core White Dwarfs.
\par
I start by summarising criteria by which pulsating stars may
be distinguished, before giving examples for the classes of
variable stars associated with the different mass ranges.
\par
It is beyond the scope of this article to discuss the
effects on the pulsations that arise when the spherical
symmetry is already broken in the equilibrium state of a
star. This happens when one orientation is preferred over
other directions, for example along the rotational axis, the
axis of a magnetic field, or the axis defined by the orbital
plane in a binary, and can have pronounced consequences
depending on the strength of the geometric, temperature or
tidal deviations of the spheroid. See the contribution by
\citet{Hekker} for a discussion on the
evolution of pulsational frequencies as a function of
magnetic fields in particular.  
\subsection{Physical seismic properties}
\textit{Geometry.\/}
In the spherically symmetric case, the solutions to the
equations describing time-dependent perturbations to the
equilibrium state separate into radial,
angle-dependent and time-dependent factors for each of the
(infinite number of) modes. Depending on the number of nodes
of the radial part, a subset of modes may be described as
low overtone (with few radial nodes, including the
fundamental mode without any radial node) or high overtone
(with many radial nodes, including the regime of 'asymptotic
limit'). The modes in a pulsator exhibiting a subset of low
overtones may also be called low-oder modes, or high-order
modes in the case of high overtones. The angular part can be
mathematically described by spherical harmonics, with the
degree describing the overall number of node planes (zero to
countably infinite) through the spherical surface, and the
azimuthal order describing the number of node planes
intersecting the equator while observing their
orientation. The azimuthal order can therefore take any
value (in integer steps) from minus the value of the degree
to plus the value of the degree. Stars with modes appearing
uniquely with a degree of zero, i.e.\ no angular node
planes, are pure radial pulsators. The occurrence of radial
and non-radial modes, i.e.\ the general case, characterizes
the non-radial pulsators. Stars in which many modes are
excited at the same time (see below) are sometimes called
multi-mode pulsators; they include non-radial modes in p-mode
pulsators and are exclusively non-radial modes
in g-mode pulsators  (see below for p and g modes) . 
\par
\textit{Restoring force.\/}
The oscillation frequency of the radial fundamental (f) mode
for a star is determined by its structure - actually almost
exclusively by the star's mean density. Although all other
modes are also global phenomena, different mode types
propagate in distinct typical depths of the star, governed
locally by oscillatory behaviour due to either predominantly
pressure or predominantly buoyancy as a restoring force.  For
both branches separately, the overall geometry of a mode,
given by its radial order, degree and azimuthal order,
determines its oscillation frequency for a given stellar
structure. If the restoring force in the structure for this
mode is predominantly conveyed via pressure (p), its
frequency will be higher than that of the radial
fundamental; if the restoring force is predominantly
conveyed via buoyancy (gravity, g), its frequency will be
lower than that of the radial fundamental. All modes with
periods longer than the fundamental radial mode are
necessarily non-radial modes.  For modes of the same degree,
a mixed mode character that bridges the evanescence zone
between the gravito- and acoustic-dominated propagation
regions in different depths can sometimes also occur.
\par
\textit{Driving mechanism.\/}
Although every star has an infinite number of such
eigenmodes, by far not all modes are excited in all stars. In
stars that are stable, all the modes are damped. This is in
fact the case for the Sun, where the oscillation caused by
any single perturbation is damped away with a characteristic
lifetime. The surface convection zone however provides
continued perturbations, or noise, at a large range of
frequencies, collectively resulting in the phenomenon of
stochastic excitation. The eigenmodes re-appear endlessly
with very low amplitudes and at random phases.
\par
A functioning driving mechanism is required to excite larger
amplitudes in at least a subset of the modes. A valve effect
in a 
suitable envelope layer of the star can turn the system into
a heat engine. In the most common case, the valve affects
the radiative flux and is provided by the opacity $\kappa$
whenever said opacity -- in contrast to its 'normal'
dependency on temperature and pressure -- increases with
compression, a behaviour found in partial ionisation zones.
The $\kappa$ effect can be further enhanced by increased
ionisation during compression, called the $\gamma$ effect,
hence also the description as $\kappa$-$\gamma$ mechanism.
\par
The partial ionisation zones of \ion{H}{} and \ion{He}{i}
give rise to $\kappa$-mechanism-driven pulsations across the
Hertzsprung-Russel diagram, ranging from Giant stars at low
effective temperatures (Mira type variables), probably
across the main sequence (roAp stars), and down to cool
DA White Dwarfs (ZZ Ceti variables). 
Where the location of the
partial ionisation zone of \ion{He}{ii} is at the right
depth in the star, we find the classical instability
strip containing the various types of Cepheids, among
them the $\delta$ Cephei, RR Lyra, and $\delta$ Scuti
variables. 
Towards higher temperatures still mainly the partial
ionisation of the M-shell electrons of \ion{Fe}{} group
elements cause the instability regions with $\beta$ Cephei
and SPB variables on the main sequence, sdBV$_\mathrm{r}$
and sdBV$_\mathrm{s}$ on the extreme horizontal branch, and
stretching up into the regime of higher luminosities where a
sdOV resides. In the region of the highest stellar
temperatures, partial ionisation of the K-shell electrons of
\ion{C}{} and \ion{O}{} drives pulsations in the
pre-White Dwarf stars of spectral type PG\,1159, the GW Vir
variables. In several of these settings, the elements
providing the opacity are supported in the 'right' stellar
depth either by radiative levitation or by mass loss that
slows down gravitational settling.
\par
The action of the $\kappa$ mechanism is modified by
convection. The pulsation-convection interaction is complex,
but when modelled successfully can explain the location of
the red edge (lower effective temperature limit) of the
instability strip. Two limiting cases, where the time scales
of convection and pulsation much differ, lead to important
further driving mechanisms in their own right. Both apply to
stellar structures with radiative interiors and exterior
convection layers. For convection time scales much longer
than the pulsation time scale, the faster luminosity
perturbations from deeper layers in the radiative flux are
'ignored' by the convective layer which is basically inert
on these time scales, and the perturbations are effectively
blocked at the base of the convection zone. This translates
into heating in phase with compression within the convection
zone, our previously formulated requirement for driving. The
convective blocking mechanism causes the excitation of modes
as seen in the $\gamma$ Dor instability region. For
convection time scales much shorter than the pulsation time
scale, surplus energy is immediately redistributed within the
convection zone, also leading to heating in phase with
compression and hence a valid driving mechanism that acts in
White Dwarfs of DBV type (V777 Her) and contributes to the
driving in the DAV type (ZZ Ceti).
\par
We complete our systematics of driving mechanisms by
mentioning the $\epsilon$ mechanism and the strange modes.
The $\epsilon$ mechanism acts on the principle of increased
heating of a layer (typically in a star's central region)
upon compression due to strongly increasing nuclear fusion
rates with increasing temperature. Changing nuclear reaction
rates during a pulsational cycle have been predicted to
affect a variety of stellar classes, but modes theoretically
excited by it will in general be unobservable as a matter of
principle due to the time scales involved. Strange modes
occur in an environment that is dominated by
radiation, i.e.\ they are relevant for stars at high
masses. The strange modes are also distinct from the p and g
modes.
\par
Different driving mechanisms tend to excite different types
of modes, e.g.\ while stochastic excitation selects
high-overtone p and g modes in solar-like stars,
the $\kappa$ mechanism due to \ion{He}{ii} excites
low-overtone p and g modes in $\delta$ Scuti stars, and 
convective blocking excites high-overtone g modes in
$\gamma$ Doradus stars. As this can sometimes happen in the
same star, the existence of hybrids is no surprise. Such
hybrids have however only been discovered in larger
numbers in the last decade (see below).
\par
\textit{Thermodynamics.\/} The discussion of mode
geometry, restoring forces and to some extent even of
driving mechanisms in a linear (where perturbations are
assumed to be small) and adiabatic approximation (where the
energy balance is assumed to be unaltered after a full
pulsational cycle) yields viable results with respect to the
prediction of the location of frequencies in a star's
pulsational spectrum. For the purpose of further
distinguishing between deviating behaviour in different
pulsators, it can be instructive to determine whether a star
adheres to linear behaviour in most of its perturbed
quantities and/or near-adiabacity in the energy conservation.
Full non-linear non-adiabatic treatment is always mandatory
to predict pulsation amplitudes, which are in reality often
not very well matched yet by models.
\subsection{Observable seismic properties: examples}
\textit{General observables.\/}
Variable stars are classified strictly by a combination of
observational properties: their pulsation periods,
amplitudes, and their spectral type and luminosity class.
Of course these characteristics are directly related to the
properties discussed above: spectral type and luminosity
class correspond to an evolutionary state that determines
the stellar structure, and the combination of driving
mechanism, mode type and selected modes in this structure
directly determines the observed pulsation periods and
amplitudes. Understanding these physical mechanisms in
detail leads to a number of useful diagnostics. Mode
identification is one possible helper in the process.
Because of geometric cancellation, only low-degree modes
retain observable amplitudes (in velocity or intensity) when
integrated across the stellar surface. This limitation of
unavoidable integration does not apply to the Sun.
\par
\textit{Solar-like oscillations.\/} Several diagnostics
have been in place well before the first solar-like
oscillations could be observed in a star other than the
Sun. The oscillations correspond to high-overtone p modes in
the envelope (the g modes believed to exist in the core are
much harder to detect, see below). This together with the
validity of the Cowling approximation, where it is
appropriate to neglect perturbations to the gravitational
potential, simplifies the equations of stellar pulsation
such that an analytic asymptotic relation for the
frequencies may be found: in this asymptotic limit,
frequencies of subsequent radial orders are evenly
spaced. These regular so-called 'large spacings' are indeed
found both in the Sun and in solar-like pulsators, and along
with the 'small spacings' found between modes of the next
but one degree give rise to such well-known diagnostics as
the {\'E}chelle diagram, the asteroseismic HR diagram (or JCD
diagram), and many more advanced tools.
\par
A major issue in applying these tools to the large number of
solar-like oscillators known today is the derivation of
proper scaling relations for observables such as the
frequency of maximum power and the mean frequency spacings,
but also the oscillation amplitudes and line widths, that
describe how these quantities vary with mass, effective
temperature, luminosity, and chemical composition of the
star. This quest for the correct scaling relations, beyond
a power of the luminosity-to-mass ratio where simply the
exponent needs to be fixed, includes the solar-like
oscillations observed in Subgiant and Red Giant stars, and
any empirical relation proposed must also be corroborated by
a theoretical understanding of it.
\par 
For the solar-like oscillations in Red Giants, a further
important diagnostic has emerged that exploits the existence
of modes with mixed character, where the information about
the core is encoded in the g mode part and communicated
through to the p mode part near the surface, which has a
detectable amplitude. In particular, the regular period
spacing of these mixed modes is readily observable.
\par
\textit{Subdwarf B stars.\/} The Subdwarf B stars may be
considered intermediate objects between solar-like stars and
White Dwarfs. With surface gravities above
$\log{g\,/\,\mathrm{cm\,s}^{-2}}\approx{5.2}$ they belong to
the compact pulsators. The propagation regions are the
envelope for the p modes and deeper regions for the g modes,
as in the Sun. The p modes, as observed in the
sdBV$_\mathrm{r}$, however are of low order and low degree.
The g modes observed in the sdBV$_\mathrm{s}$ are of mid
orders and, as demonstrated by \citet{2000ApJS..131..223C},
are similarly trapped and confined as the g modes in
pulsating White Dwarfs. Due to the low or mid overtones,
equal frequency spacings for the p modes or equal period
spacing for the g modes in the asymptotic limit may be
approached in some stars but are not strictly realised, so
that full modelling is always required do to asteroseismology
with sdB variables. In particular, there is no physical
reason for the existence of, and no observational indication
for, small frequency spacings \citep{2007ApJ...654.1087F}.
\par
\textit{White Dwarfs.\/}
The frequencies observed in White Dwarfs correspond to
high-order low-degree g modes in GW Vir variables, and to
low-order low-degree g modes in ZZ Ceti variables. These
modes, propagating in the envelope, are attributed to the g
branch although the restoring force is a combination of
acoustic and gravity effects, which is equally true for the
p modes found in the core of White Dwarf models. The latter
are not only hard to detect because of their near-vanishing
amplitudes at the surface, but also due to their very short
periods of the order of one second, and have not been
observationally discovered yet. This is quite the opposite
from the situation in the Sun, where it is the g modes that
live in the core and that are observationally elusive.
\par
In the case of the GW Vir variables, the regular period
spacing of the high-order g modes can be explored as
diagnostics. For selected GW Vir stars,
\citet{2009A&A...499..257C} have systematically investigated
the differences in their results obtained for the stellar
parameters by comparing the observed period spacing with the
\emph{asymptotic} period spacing of their models, and with
the \emph{average} of the computed period spacings, as well
as with their results that they obtained through a direct
fit procedure to individual periods.  
\section[High-mass stars]{High-mass stars}
Stars at high masses are intrinsically highly variable. The large radiation 
pressure leads to dynamic atmospheres with significant mass loss. The light
observed from such stars therefore often appears modified by the previously
ejected circumstellar material that it passes through. The various physical
processes are not always well enough understood to clearly disentangle the
origin of an observed variation, e.g.\ from rotation or pulsations.
\subsection{Vicinity of the main sequence}
Pulsations in B stars are driven by the \ion{Fe}{}-$\kappa$ mechanism. At two
to seven solar masses, i.e.\ intermediate masses, the Slowly Pulsating B (SPB)
stars pulsate in high-order g modes. In the seven to twenty solar mass range,
the more rapid $\beta$ Cephei variability results from low-order p and g
modes. Some Be stars are also pulsating and closely linked to the SPB and
$\beta$ Cep. Additionally, (very few) rapidly rotating SPB stars exist with
their frequency spectra modified accordingly. Making things yet more
complicated, slow pulsations are also detected in all $\beta$ Cephei stars as
soon as the sensitivity is high enough, i.e.\ when the objects are observed
from space (see also the discussion on hybrid pulsators). At higher masses,
the formerly separately classified $\zeta$ Oph variables of
spectral type O have been recognised as the high-temperature
extension of the $\beta$ Cep variables. In this regime, strange mode driving
due to the increased radiation pressure starts to add to the 
\ion{Fe}{}-$\kappa$ mechanism and eventually becomes dominant. The
theoretical instability region widens and starts to encompass the main
sequence above forty solar masses, as well as the supergiant region at low
effective temperatures.
\subsection{Evolved stages}
At later evolutionary stages, the mass loss suffered previously leads to a
higher luminosity-to-mass ratio and hence an increasing contribution from
strange modes. Above twenty solar masses, a combination of \ion{Fe}{}-$\kappa$
mechanism and strange modes, with unclear amounts of contributions from each,
generates more or less irregular variability in stars. The boundaries between
pulsations and outbursts start to dissolve.  The Periodically Variable
Supergiants (PVSG) of spectral types A and B tend towards pulsational
characteristics, while the Luminous Blue Variables (LBV) definitely tend
towards the outburst characteristics. The locations of LBV stars in the HR
diagram may be appreciated in Fig.~4 of \citet{2012Natur.482..375R}. They are
believed to represent an intermediate evolutionary stage in the transition to
the Wolf-Rayet stars (see Fig.~3 in \citealt{1989ASSL..157..229M} for an
evolutionary sequence including the so-called Wolf-Rayet funnel in the HR
diagram). Strange modes have been predicted in many types of massive stars,
but are expected to have the largest amplitudes in Wolf-Rayet stars. The
detection of a periodic signal of potentially pulsational nature in a
Wolf-Rayet star observed by the MOST satellite however exhibits an
unexpectedly long period \citep{2005ApJ...634L.109L}. Although this and
similar observations in two further objects are attributed to pulsational
instabilities, they are not generally accepted as unambiguous strange mode
detections. 
\par
Cepheid variability also occurs partly in the high-mass
range, spanning two to twenty
solar masses for the classical Cepheids,
but is entirely discussed in the next section.
\section{Intermediate- and low-mass stars}
\subsection{The classical instability strip}
The Cepheid instability strip encompasses a wide range of
evolutionary states and masses. 
\par
\textit{Pre-main sequence stars.\/}
Pre-main sequence stars at masses above eight solar masses
are not observed as they evolve too quickly. The pre-main
sequence stars in the mass range $2$-$8$ solar masses are
the Herbig Ae/Be stars, those with masses below two solar
masses the T Tauri stars. Both
cross the instability region on their way to
the main sequence and can become pulsationally unstable,
resulting in $\delta$ Scuti-like pulsators that populate the
same area in the HR diagram as the classical $\delta$ Scuti
stars \citep{2009CoAst.159...59Z}. Some T Tauri stars are
known to additionally pulsate in high-order g modes.
\par
Apart from this passage through the classical instability
strip, it should be noted that \citet{2005JApA...26..171S}
have predicted solar-like oscillations in pre-main sequence
stars from theory. Solar-like oscillations in their own
right as well as the evolution of seismic properties of
pre-main sequence -- main-sequence -- Red Giant stars are
furthermore treated in many of the other contributions to
this issue. 
\par
\textit{Cepheids.\/}
The Cepheids encompass the classical Cepheid variables of
$\delta$ Cephei type with masses above three solar masses
(and their population II counterparts of W Virgines type
with masses below one solar mass); the horizontal-branch
stars of RR Lyrae type \citep[see also][]{Guggenberger} in
the mass range of $0.6$-$0.8$ solar masses; and the Dwarf
Cepheids of $\delta$ Scuti type with masses from $1.5$-$2.5$
solar masses (and their population II counterparts of SX
Phoenicis type, which are typically blue stragglers and
whose origin is not always certain for a given object),
found near the main sequence but actually representing
evolutionary inhomogeneous classes as described above. An
additional type are the anomalous Cepheids, believed to
result from binary mergers.
\par
Historically, one of the great mysteries in Cepheid research
has been the mass discrepancy problem, paraphrasing the fact
that masses derived from evolutionary models on the one hand
and pulsational models on the other hand were in
disagreement by up to $30$\%-$40$\%, until the usage of Los
Alamos opacities was replaced by OPAL opacities
\citep{1992ApJS...79..507R}, reducing the discrepancy to
below $10$\% \citep{1992ApJ...385..685M}. Systematic
differences still persist at this level, as shown for
example by \citet{2010Natur.468..542P} who were able to
confirm the pulsational mass of a Cepheid in an eclipsing
binary dynamically with a precision of $1$\%. The
evolutionary models therefore require improvement and at the
same time demonstrate the continuing need for better
opacities along with the re-assessment of further input
physics such as convection. The need for work on the
opacities is a major issue encountered in the analysis of
several further types of variables, including both the
pulsating main sequence B and the Subdwarf B stars,
and last but not least the Sun itself.
\par
\textit{Radial and non-radial pulsations in Cepheids.\/}
The radial pulsations in classical Cepheids come in several
flavours: The most common occurrence are single modes, with
the star pulsating either in the fundamental (F) or the
first overtone mode (O1), but the  OGLE survey in particular
has added significantly to the zoo. Still in single mode,
some Cepheids pulsate in the second overtone (O2), whereas
double mode pulsators can show both the combination F/O1 or
O1/O2, and then there are some triple mode pulsators as well
as Cepheids with Blazhko-like modulations. (For the more
common Blazhko effect in RR Lyrae stars, see e.g.\
\citealt{2010MNRAS.409.1244S}, 
\citealt{Molnar},
\citealt{Guggenberger}).
\par
Furthermore, non-radial modes have been observed by
\citet{2008AcA....58..163S} and
\citet{2009MNRAS.394.1649M}. Interestingly, these observed
non-radial pulsations cannot be explained theoretically by
linear instabilities for low degrees
\citep{2007A&A...465..937M}, so significant challenges
remain even for the poster child variable stars that are the
Cepheids. The precise knowledge of Cepheid parameters is of
paramount interest due to their role as distance
indicators. Asteroseismologically, however, Cepheids are not
as profitable as other pulsators due to the overall sparsity
of their modes: As a rule, more, and different, modes are
better to differentially sound the inner structure of stars.
\subsection{Hybrid pulsators}
In trying to meet the need to obtain improved local
and global stellar parameters, hybrid pulsators have the
potential to contribute significantly to breakthroughs.
\par
\textit{Hybrids with solar-like oscillations.\/}
The Sun itself is a prime example for a hybrid, where the
envelope acoustic modes are measured, and the quest for
gravity modes from the core is considered a major
objective. The detection of g modes via a periodic structure
in agreement with the predicted period separation for dipole
modes has been claimed by \citet{2007Sci...316.1591G}, but
\citet{2010A&ARv..18..197A} can only conclude that ''there
is indeed a consensus [\ldots] that there is currently no
undisputed detection of solar g modes''.  
\par
The situation is different for Red Giants, where all modes
are mixed modes and carry information from the stellar core.
Such mixed modes \citep{2010ApJ...713L.176B}, as well as
g-mode period spacings \citep{2011Sci...332..205B}, are
reported from observations. 
\citet{2012Natur.481...55B} use an inverse approach to
obtain results implying a fast rotation of the core. 
Forward modelling by 
\citet{2012arXiv1209.5621O} 
on the other hand suggests that, already at moderate
rotation rates, symmetrical patterns arise around
axi-symmetric modes that may be mistaken as multiplet
splittings. She further cautions that at high core rotation
rates the splittings are so far from symmetrical that the
correct selection of modes to measure splittings can become
very difficult, if not impossible.
\par
Back on the main sequence, \citet{2011Natur.477..570A}
report the detection of solar-like oscillations in a
$\delta$ Scuti star, well above the temperature limit
empirically determined to mark the transition between
opacity-driven and solar-like pulsations by
\citet{2011ApJ...743..143H} with some small overlap between
the two. The existence of the $\delta$ Scuti - solar-like
hybrid does not necessarily imply that the location of this
transition must be revised. It could also be possible that
the long lifetimes are due to energy transfer from the
$\kappa$ mechanism (Antoci, priv.\ comm.).
\par
Further up the main sequence, hybrid behaviour has
been predicted for the $\beta$ Cep and the SPB stars
\citep{2009CoAst.158..269B}. Solar-like oscillations in a
$\beta$ Cep star have subsequently been reported by
\citet{2009Sci...324.1540B}, but no further such objects
turned up in the KEPLER data investigated by
\citet{2011MNRAS.413.2403B}.
\par
\textit{Hybrids on the main sequence.\/}
The $\delta$ Scuti variables show low-order g and p modes,
the $\gamma$ Dor variables show high-order g modes. In the
overlap region between the two instability strips, four
hybrids had been confirmed prior to KEPLER, e.g.\ with
MOST; see for example
\citet{2002MNRAS.333..251H} and 
\citet{2006CoAst.148...34R}, and 
references therein.
Several of these hybrids show intermediate-order gravity
modes, challenging theory. The difficulties have only
increased so far since the numbers of known hybrids has
gone up significantly with detections from CoRoT and KEPLER,
across a parameter space that stretches beyond the bounds of
both previously established instability regions. The
situation to date is that there exist no pure $\delta$ Sct
or $\gamma$ Dor regions. The hybrids detected from space fill
the $\delta$ Sct instability strip and do not simply
concentrate on the overlap region \citep[cf.\ ][]{Hareter}.
\citet{2011A&A...534A.125U} find that 23\% of 171 A- and
F-type stars are hybrids. While \citet{2010ApJ...713L.192G}
go ahead and boldly propose a new classification scheme,
\citet{2011A&A...534A.125U} stress the importance of first
determining accurate physical parameters of all the stars.
They continue by suggesting that a physical mechanism
different from the $\kappa$ mechanism and convective
blocking effects might be responsible for the occurrence of
hybrids well outside the $\gamma$ Dor instability region
that must be investigated.
\par
Further peculiar combinations include the detection of
$\gamma$ Dor pulsations in a roAP star, and $\gamma$ Dor and
$\delta$ Scu pulsations in Ap stars
\citep{2011MNRAS.410..517B}.
\par
Beyond the main sequence A-F stars, a strange mode has been
reported in a RR Lyrae star \citep{2010MNRAS.409.1244S}.
\par
A further important group are the hybrids among B stars
showing $\beta$ Cep and SPB pulsations, i.e.\ low-order p
and g modes simultaneously with high-order g modes
\citep[see
  e.g.\ ][]{2005MNRAS.360..619J,2011MNRAS.413.2403B}. 
These hybrids, however, do mostly not occur in the
overlap region, but in hotter $\beta$ Cep that exhibit 
additional high-order g modes. The task ahead for
asteroseismology is to understand and take advantage of all
these newly discovered objects. 
\par
\textit{Hybrids among compact pulsators.\/}
The compact pulsators comprise evolved objects either beyond the Red Giant
Branch or post-Asymptotic Giant Branch phases. On their way to exposure of
the compact core, stars evolving through the AGB stage may first become Mira
variables. As post-AGB stars, they may turn into RV Tauri variables, which are
Supergiants pulsating in the fundamental mode. Hydrogen-deficient post-AGB
stars are discussed at the end of this subsection. Hot Subdwarf stars are
bound to completely avoid the AGB phase.
\par
The Subdwarf B stars show either rapid (sdBV$_\mathrm{r}$)
or slow (sdBV$_\mathrm{s}$) pulsations, or both: the first
hybrids (sdBV$_\mathrm{rs}$) were found from the ground in
the empirically determined short-period/long-period overlap
temperature range
\citep[e.g.\ ][]{2006A&A...445L..31S}. From KEPLER data,
\citet{2010MNRAS.409.1470O,2011MNRAS.414.2860O} find that
below the temperature boundary region marked by the hybrid
sdB pulsators discovered from the ground, all stars pulsate
when observed from space.  Not unlike the situation for
$\beta$ Cep stars when observed from space, where SPB
pulsations are always present as soon as the sensitivity is
high enough, \citet{2010MNRAS.409.1470O} demonstrate that
the only rapid pulsator in the KEPLER field also shows a
low-amplitude mode in the period region of slow
pulsations. Moreover, from the additional detection of
rapid pulsations with low amplitudes in slowly pulsating sdB
stars, located in the lower-temperature parameter
range, they conclude that hybrid behaviour may be common in
these stars even outside the boundary temperature region
where hybrid pulsators had previously been found.
\par
A very peculiar object among the pulsating Subdwarf B stars
is the He-sdB star LS IV$-$14$^{\circ}$116. When
\citet{2005A&A...437L..51A} discovered its pulsations, they
attributed them to high-order g modes. This has been
confirmed by model calculations by
\citet{2011ApJ...741L...3M}, with an interesting twist: in
their models, the $\epsilon$ mechanism acting in the
\ion{He}{}-burning shells is driving the
pulsations. If confirmed, this would be the first
'discovery' of the $\epsilon$ mechanism in a star that
corresponds to an object actually observed in reality. 
\par
For completeness, I also mention the Subdwarf O stars where
new classes of variables might be emerging. Only one
variable sdO is known in the field
\citep{2006MNRAS.371.1497W}, and it shows very rapid
multi-frequency variations.  \citet{2012ASPC..452..241R}
report on a group of sdOV, with seismic properties closer to
those of the field sdB variables, in the cluster $\omega$
Cen.
\par
Within the White Dwarf regime, all the instability regions
are well separated from each other. While there are hence no
hybrids, White Dwarfs have repeatedly been the targets for
\smash{$\dot{P}$} measurements which are discussed in the
following subsection. The pulsating White Dwarf classes are
therefore introduced here.
\par 
The majority of White Dwarfs, the DAs with
hydrogen-dominated atmospheres, move through the ZZ Ceti
instability strip furthest down on the cooling sequence.
\par
The hottest pulsating White Dwarfs, the GW Vir variables,
belong to the two (evolutionary probably subsequent)
spectral classes Wolf-Rayet-type Central Star of Planetary
Nebula ([WC]) and PG\,1159, which are believed to emerge
from a second passage to the Asymptotic Giant Branch after a
(late) helium flash, known as the born-again scenario. This
significantly alters their surface composition: Notably, the
atmospheres become hydrogen-deficient, with individual
differences in the abundances of the dominant elements
\ion{He}{}, \ion{C}{}, \ion{O}{} and \ion{Ne}{}. The
descendants of the [WC] to PG\,1159 sequence are the
helium-rich DO, and later DB, spectral type objects, unless
there were residual traces of hydrogen left in the
PG\,1159. There are no stars spectroscopically classified as
DO among the GW Vir variables, so the alternative
designation DOV is misleading.
\par
The White Dwarfs with helium-dominated atmospheres of DB
spectral type eventually make their passage through the DBV,
or V777 Her, instability strip.
\par
The wide variety of the hydrogen-deficient (pre-)White
Dwarfs and their possible evolutionary links have been
described by \citet{2011arXiv1109.2391W}. Among them are the
Hot DQ stars, from which stems the most recently introduced
class of pulsating White Dwarfs, the Hot DQ pulsators
\citep{2008ApJ...678L..51M}. A significant fraction of them,
currently five out of a total of fourteen of these rare
objects, pulsate in low-order and low-degree g modes
\citep[][and references therein]{2011ApJ...733L..19D}.
Following latest suggestions for their evolutionary origin
\citep{2011arXiv1109.2391W}, the Hot DQ stars may be the
possible descendants of super-AGB stars, and would not have
evolved through the 'normal' PG\,1159 stage as previously
considered.
\par
In this context, it is also opportune to refer to a potential additional
hydrogen-deficient post-AGB evolutionary sequence, encompassing further
variable classes. If the R\,CrB stars are the result of a binary merger of a
\ion{He}{}- and a \ion{C}{}/\ion{O}{}-core White Dwarfs (instead of descending
from a born-again track after a late thermal pulse), then they are probably
also the progenitors of the extreme helium stars (eHe), which subsequently
evolve into \ion{O}{}(\ion{He}{}) White Dwarfs. The R\,CrB phenomenon is
characterised by a very noticeable variable behaviour consisting of irregular,
sudden dimmings by up to several magnitudes, believed to result from dust
formation episodes, followed by gradual recovery to the previous brightness.
The R\,CrB stars have spectra of types F and G, the eHe stars spectra of type
A and B. At lower luminosities corresponding to Giant stars, the
hydrogen-deficient carbon stars (HdC) show very similar characteristics as the
R\,CrB stars, minus the R\,CrB variability. All of these stars have relatively
low masses below one solar mass, mostly concentrated in a compact core, but
surrounded by an extended envelope that leads to a very high overall
luminosity-to-mass ratio. Just as for the massive main sequence or evolved
stars, this high luminosity-to-mass ratio leads to pulsational instabilities
that are observed, at lower amplitudes than the R\,CrB phenomenon itself, in
practically all R\,CrB stars and related objects. The contribution of strange
modes is significant, in particular for the g-type or radial modes in the
R\,CrB stars, and for the longest-period modes, as well as the short-period
non-radial g modes, in the eHe stars. Finally, in between those period ranges,
a third variable class is known among the eHe stars that seems to be better
characterised as g modes driven by the \ion{Fe}{}-$\kappa$ mechanism.
\subsection{Secular \smash{$\dot{P}$} effects}
Period changes in pulsating stars are a common phenomenon,
and the Cepheids may again serve as a first example here. A
non-zero \smash{$\dot{P}$} has been reported for RR Lyrae as
early as \citeyear{1916ApJ....43..217S} by
\citeauthor{1916ApJ....43..217S}. The potential of such
measurements to constrain stellar evolution scenarios was
highlighted by \citet{1919Obs....42..338E}, when he
concluded from the small value of the period shortening
observed for $\delta$ Cephei that it was quite inconsistent
with the much larger value ''demanded by the contraction
theory'', leaving him ''no escape from the conclusion that
the energy radiated by a star comes mainly from some source
other than contraction''. Later, \citet{1959S&T....18..309S}
worked out how the period change in Cepheids is due to a
change in radius, with increasing or decreasing effective
temperature at constant luminosity. The sign of the period
change therefore determines the instability crossing mode
(from higher to lower temperatures or vice versa), and its
absolute value can indicate whether it is the first, second,
or third pass, as \citet{2006PASP..118..410T} have recently
shown by comparing the observed and theoretical
\smash{$\dot{P}$} for a sample of 200 Cepheids.
Figuratively, obtaining a \smash{$\dot{P}$} value puts an
arrow of defined length and at least with a primer on the
direction (negative or positive) onto an existing evolutionary
track at the current location of the object, with the
evolutionary track simultaneously containing its on
theoretical values for \smash{$\dot{P}$}.
Many of them could build up something similar to a slope
field for evolutionary tracks.  
\par 
Assuming a detailed model for a pulsating star exists from
asteroseismology, one \smash{$\dot{P}$} measurement -- or
better yet more for different modes -- can provide an
excellent test of the quality of the underlying evolutionary
model and potentially constrain it even further. Such
measurements are observationally expensive and, depending on
the type of star, may require decades of monitoring for a
conclusive result. The measured values available with
components that actually represent true secular effects are
therefore rather rare. They imply that we are following
stellar evolution in real time; here are some examples. 
\par
For the Subdwarf B stars, a monitoring program is on-going
\citep{2010Ap&SS.329..231S}, and results are available for
three objects so far (\citealt{2002A&A...389..180S},
\citealt{2011PhDT........85L}), including preliminary
evidence on the degree of their core helium exhaustion.
\par
For the prototype of the GW Vir variables,
\citet{2008A&A...489.1225C} present measurements that
illustrate how complex and at the same time rewarding the
interpretation of \smash{$\dot{P}$} can be. In this region,
the PG\,1159 stars turn around the 'knee' in a
$\log{g}$-$T_{\mathrm{eff}}$ diagram, and can take on both
negative and positive values for \smash{$\dot{P}$}.
\par
The \smash{$\dot{P}$} values are always positive further
down the cooling sequence of White Dwarfs, and since the
cooling rate has gone down significantly once the ZZ Ceti
instability strip is reached, the values become smaller and
harder to measure. \citet{2005ApJ...634.1311K} have
succeeded in obtaining a significant measurement for one
DAV, using it to derive the \ion{C}{}/\ion{O}{} ratio in
the core. \citet{2008ApJ...676..573M} have extended the
survey to a dozen further objects.
\section[Very-low-mass stars and beyond]{Very-low-mass stars and beyond}
\subsection{M Dwarfs}
%
In the centre of stars below $0.5$ solar masses, the
conditions for stable core helium fusion are never reached,
so these stars will eventually evolve into He-core white
dwarfs. Stars become fully convective below $0.25$ solar
masses and never evolve into Red Giants. At lifetimes of
$5\cdot10^{11}$ years and longer on the main sequence, well
beyond the present age of the universe, evolutionary
considerations are however of no practical concern for the
discussion of presently known classes of pulsating stars in
this mass range.
\par
Observationally, M dwarfs appear as quite variable stars due
to their significant levels of activity. This intrinsic
variability in conjunction with their relative faintness
hampers the search for the signature of pulsations, and no
detections have been reported so far.
\par
In principle, solar-like oscillations are expected in any
star with a surface convection zone, so on the main sequence
this applies for all stars from one solar mass down,
including the M dwarfs discussed here. The stochastic
excitation should be even more effective once the stars
become fully convective, but at the same time (in)stability
against pulsations also becomes increasingly harder to
compute. Although \citet{2011ApJ...743..143H} report some
evidence for the hypothesis that increased stellar activity
suppresses mode amplitudes via the magnetic fields (albeit
predominantly for Subgiant stars), it should only be a
question of sensitivity to eventually detect the
stochastically excited oscillations, and the first
radial-velocity searches are underway.
\par
Additionally, as previously for many other types of stars,
it has been proposed from theoretical calculations that
pulsations may be excited in M Dwarfs through the $\epsilon$
mechanism. As in previous investigations, however, it turns
out that the growth rates, describing the time after which
the modes reach an observable amplitude, are of the same
order of magnitude as the time scales of stellar evolution,
i.e.\ the nuclear time scale.  These modes are therefore
unobservable as a matter of principle.
\subsection{Brown Dwarfs}
Beyond the bottom of the main sequence, the Brown Dwarfs are
completely uncharted territory with respect to
pulsations. The difficulties are on the one hand of
observational nature -- the objects are much fainter still
than the M Dwarfs -- but it is also theoretically difficult
to explore this mass range. At the temperatures and
pressures associated with the structures of Brown Dwarfs,
the uncertainties in the opacities, in particular the
molecular opacities, are currently still so large that the
results of any stability analysis cannot be considered very
reliable.
\subsection{Giant Planets}
What is true for Brown Dwarfs when it comes to observational
difficulties holds even more with respect to planets in
general, but the planets in our own Solar System are an
exception. In particular, global oscillations have been
predicted and searched for in the Gas Giant Jupiter for
quite some time, and it looks like they have finally been
securely detected last year by
\citet{2011A&A...531A.104G}. Interestingly, the SYMPA
instrument, built for and used in that work, has been
modelled on similar instruments (GONG and MDI/SOHO) dedicated
to solar observations, with observations resulting in a
two-dimensional radial velocity map. In this sense this
result builds a bridges from our discussion of
asteroseismology back to the complementary topic of
helioseismology.
\section{Summary}
Variable stars are classified by their periods and
amplitudes. In order to fully characterise the pulsational
behaviour when the variations are intrinsic, it is necessary
to explore properties such as driving mechanism, character
of the restoring force, and mode geometry along with the
structure (given by the mass and evolutionary state) of the
star. For several classes of pulsators, a good understanding
of the seismic properties has lead to the development of
useful advanced diagnostic tools, several of which have been
mentioned here for reference.
Among the examples given for recent research results is the
finding that the Hertzsprung-Russel diagram is teeming with
asteroseismic hybrids, which have been described as
completely as possible, that await to be explored. 
Beyond the classical asteroseismology exercise of matching
observed periods to those in a set of models, thus deriving
stellar parameters, $\dot{P}$ measurements have been
recalled to provide an additional test for consistency
between evolutionary and pulsational models.
The observational difficulties as well as challenges
encountered in the interpretation of observations 
have been showcased across a wide range of stellar masses
and ages.
\acknowledgements
The author would like to thank the organisers, including the ESF staff, for a
very nice conference which provided a wonderful occasion for interesting
discussions. Among the exchange with colleagues, the input by Victoria Antoci,
Gilles Fontaine, Wolfgang Glatzel, Gerald Handler, Markus Hareter, and
Rhita-Maria Ouazzani has been especially helpful in preparing this article.
The author's invited participation, including full support for the
accompanying babysitter, was generously funded by the conference organisers
through their sponsors ESF (European Science Foundation), HELAS (The European
Helio- and Asteroseismology Network) and EAST (The European Association for
Solar Telescopes).


\end{document}